\documentclass[12pt]{iopart}
\usepackage{iopams}
\usepackage{setstack}
\usepackage{amsthm}
\usepackage{amstext}
\usepackage{amsfonts}
\usepackage{amssymb}
\usepackage{graphicx}
\usepackage{subfigure}
\usepackage{array}
\usepackage{appendix}
\usepackage{mathrsfs}
\usepackage{color}
\usepackage{xcolor}
\usepackage{makecell}
\usepackage{adjustbox,lipsum}
\usepackage{bibentry}
\usepackage{braket}
\usepackage{cite}

\begin{document}

\title{Photonic variational quantum eigensolver using entanglement measurements}

\author{Jinil Lee$^{1,2}$, Wooyeong Song$^{1,3}$, Donghwa Lee$^{1}$, Yosep Kim$^{1,4}$, Seung-Woo Lee$^{1}$, Hyang-Tag Lim$^{1,2}$, Hojoong Jung$^{1}$, Sang-Wook Han$^{1,2}$, and Yong-Su Kim$^{1,2}$}

\address{$^1$ Center for Quantum Information, Korea Institute of Science and Technology (KIST), Seoul, 02792, Republic of Korea}
\address{$^2$ Division of Quantum Information, KIST School, Korea University of Science and Technology, Seoul 02792, Republic of Korea}
\address{$^3$ Quantum Network Research Center, Korea Institute of Science and Technology Information (KISTI), Daejeon 34141, Republic of Korea}
\address{$^4$ Department of Physics, Korea university, Seoul, 02841, Republic of Korea}

\begin{indented}
    \item{The first two authors (J. L. and W. S.) contributed equally to this work and can be regarded as main authors.}
\end{indented}

\ead{yong-su.kim@kist.re.kr}

\begin{abstract}
Variational quantum eigensolver (VQE), which combines quantum systems with classical computational power, has been arisen as a promising candidate for near-term quantum computing applications. However, the experimental resources such as the number of measurements to implement VQE rapidly increases as the Hamiltonian problem size grows. Applying entanglement measurements to reduce the number of measurement setups has been proposed to address this issue, however, entanglement measurements themselves can introduce additional resource demands. Here, we apply entanglement measurements to the photonic VQE utilizing polarization and path degrees of freedom of a single-photon. In our photonic VQE, entanglement measurements can be deterministically implemented using linear optics, so it takes full advantage of introducing entanglement measurements without additional experimental demands. Moreover, we show that such a setup can mitigate errors in measurement apparatus for a certain Hamiltonian.
\end{abstract}

%
%

\section{Introduction}
Quantum computing technology has been rapidly developing, and it opens so called noisy intermediate-scale quantum (NISQ) era~\cite{preskill2018}. One of the major tasks in the NISQ era would be finding practical applications of quantum computers which have a limited number of qubits and non-negligible errors. Variational quantum algorithms (VQA) have attracted a lot of interest as promising application candidates for NISQ devices~\cite{farhi2014, Yuan2019, Bravo-Prieto2019, lubasch2020, liang2020, barron2020, cerezo2021, xu2021, benedetti2021}. VQAs are classical-quantum hybrid algorithms that efficiently represent the problems using quantum systems and utilize a classical computer to effectively optimize the quantum systems. 

Variational quantum eigensolver (VQE) is one of the most spotlighted example, which solves eigenvalue problems~\cite{peruzzo2014}. It has several notable advantages in NISQ devices. First, the necessary measurements for estimating expectation value of a given Hamiltonian are expressed as a linear combination of single-qubit measurements, making it relatively easy to implement quantum circuits. Second, the computations performed through the optimization of parameterized quantum circuit are implemented by well-developed classical algorithms. Finally, it aims at addressing highly practical problems such as finding the ground state and its energy of a molecular Hamiltonian. Since the seminar work by Peruzzo {\it et al}.~\cite{peruzzo2014}, it has been implemented various physical platforms including superconducting qubits, ion traps and photonic qubits~\cite{o2016,kandala2017,kokail2019,Borzenkova2021,lee2022,zhao2023}.

Despite these notable advantages, there are still many obstacles when it comes to practical-level applications~\cite{mcclean2018, bittel2021}. One important issue is the resource requirement for state preparation and measurement (SPAM). Specifically, in the ordinary VQE, it is required to prepare the trial state and perform a large number of Pauli measurements representing the problem Hamiltonian. However, as the size of Hilbert space that the Hamiltonian deals with increases, the number of required projective measurements and parameters to be optimized drastically increases~\cite{bittel2021}.
As one of the proposals to solve such a challenge, utilizing entanglement measurements to reduce the number of measurement setups has been proposed~\cite{Gokhale2019,hamamura2020,kondo2021, escudero2022}. In specific, by extending the commutativity of measurement operators with entanglement measurements, it is possible to reduce the number of measurement setups through simultaneous measurements. However, entanglement measurements require additional two-qubit operations, and thus, the possible advantage can be dimmed as the two-qubit operation error increases. 

Recently, experimental resource-efficient photonic VQE, that simultaneously utilizes polarization and path degrees of freedom of a single-photon, has been implemented~\cite{lee2022}. One advantage of this encoding method is that the CNOT operation can be deterministically implemented using linear optical elements. Therefore, this photonic VQE can leverage entanglement measurements regardless of the experimental burdens associated with additional two-qubit operations. In this paper, we experimentally explore a photonic VQE with entanglement measurements, and explicitly show that the number of measurement setups can be reduced by applying entanglement measurements. We also demonstrate that the ground state energy estimated by VQE with entanglement measurements can be robust against operational errors for a certain Hamiltonian. 


\section{Theory}
\subsection{Variational Quantum Eigensolver}
VQE is a quantum-classical hybrid algorithm to obtain the eigenvalues of matrices such as Hamiltonian operators. VQE iteratively minimizes the Rayleigh quotient $R ( H, \ket{\psi} )$ to find a minimum energy eigenvalue by varying the quantum state $\ket{\psi}$.
\begin{eqnarray}
R(H,\ket{\psi}) = \frac{\braket{\psi|H|\psi}}{\braket{\psi|\psi}} = \braket{H}_\psi.
\end{eqnarray}
Here, $\braket{\psi | \psi} = 1$, and the Rayleigh quotient becomes an expectation value of Hamiltonian $H$, $R(H, \ket{\psi} )= \braket{H}_{\psi}$. Since the ground-state energy $E_g$ is the minimum energy, it should always be less than or equal to the expectation value of Hamiltonian, $E_g\leq\braket{H}_{\psi}$. 

In order to obtain the expectation value $\braket{H}_{\psi}$, we express Hamiltonian using Pauli operators to be more appropriate to be dealt with qubits,
\begin{eqnarray}
H = \sum_{j} w_j {\sigma}_j,
\end{eqnarray}
where $ w_j \in \mathbb{R} $, and $ \sigma_j $ are the Pauli strings that are tensor products of Pauli operators. In other words, a Hamiltonian can be represented in a weighted sum of Pauli strings. Then, the expectation value of Hamiltonian can be determined by obtaining the individual expectation values of Pauli strings as
\begin{eqnarray}
\braket{H}_{\psi} = \sum_j w_j \braket{\sigma_j}_{\psi}.
\label{eq_H_exp}
\end{eqnarray}

The quantum state $|\psi\rangle$ is parameterized by controllable parameters to be optimized, and it serves as the trial state or `ansatz'. Then, we obtain the expectation value of the Pauli string by applying projective measurements of the state on the bases represented by the Pauli string. By performing this procedure for all Pauli strings which constitute the Hamiltonian, the expectation value of Hamiltonian can be constructed by adding them all together with their weights $w_i$. Then, algorithmic processes based on classical computers optimize the parameters of the ansatz quantum state to reduce the expectation value. After iterative procedures, we can reach the minimum value, which is the ground state energy, and its corresponding ansatz state, i.e., the ground state. Note that the optimization process is performed by a classical computer, so we can utilize well-developed classical optimization methods.

\subsection{Grouping measurements}

In VQE, the role of quantum processing unit (QPU) is to prepare an arbitrary ansatz state and to perform projective measurements to find the expectation value of the Pauli strings. In conventional VQE, single-qubit measurements are usually performed since they are easy to realize. In practice, this is one of the strong points of VQE. However, as the quantum system and Hamiltonian becomes larger and more complicated, the number of Pauli strings can rapidly increase. Thus, it becomes challenging to implement in a NISQ QPU. One way to relax this complication is to group Pauli strings that are simultaneously measurable and reduce the number of measurement setups~\cite{Gokhale2019}. For example, $XY$, $XI$, and $IY$ are `qubit-wise' commutative and simultaneously measurable using a single-qubit measurement setup.  

The Pauli string grouping can be more efficient by considering more general commutativity. For instance, $XX$ and $YY$ do not commute in terms of qubit-wise commutativity, i.e., $X$ and $Y$ are not commute. However, they commute in terms of general commutativity considering the overall string. Therefore, they can be simultaneously measured by a single measurement setup. This can be achieved by introducing more general measurements such as entanglement measurements~\cite{Gokhale2019, hamamura2020, kondo2021, escudero2022}. However, entanglement measurements require additional two-qubit operations which can be noisy in NISQ machines. Therefore, the benefit-and-cost of introducing entanglement measurements in VQE experiment should be carefully considered.

As an example, we estimate the ground state energy of two-qubit antiferromagnetic Heisenberg model, $H = XX+YY+ZZ$. This Hamiltonian vividly shows reducing the measurement setups by applying entanglement measurement. In conventional Pauli measurements, three measurement setups which project qubits on to $XX$, $YY$ and $ZZ$ are required. On the other hand, these three Pauli strings are simultaneously measurable by introducing an entanglement measurement. For example, if one performs a projective measurement on Bell states, $|\psi^{\pm}\rangle=\frac{1}{\sqrt{2}}\left(|01\rangle\pm|10\rangle\right)$ and $|\phi^{\pm}\rangle=\frac{1}{\sqrt{2}}\left(|00\rangle\pm|11\rangle\right)$, the Pauli strings can be reconstructed as follows:
\begin{eqnarray}
XX &=& |\psi^+\rangle \langle \psi^+| + |\phi^+ \rangle \langle \phi^+| - |\psi^- \rangle \langle \psi^- | - |\phi^- \rangle \langle \phi^-|, \nonumber \\
YY &=& |\psi^+\rangle \langle \psi^+| + |\phi^- \rangle \langle \phi^-| - |\psi^- \rangle \langle \psi^-| - |\phi^+ \rangle \langle \phi^+ |,  \\
ZZ &=& |\phi^+\rangle \langle \phi^+| + |\phi^- \rangle \langle \phi^-| - |\psi^+ \rangle \langle \psi^+ | - |\psi^- \rangle \langle \psi^-| \nonumber.
\label{byBSM} 
\end{eqnarray}
This shows that they can be represented as the sum of projections on Bell bases. Therefore, the VQE implementation of Hamiltonian $H$ can be obtained from a single measurement setup. 

As an additional example, we also showcase the HeH$^+$ cation Hamiltonian~\cite{lee2022}. The number of measurement setups in this Hamiltonian is reduced from four to three by introducing entanglement measurements. The full Hamiltonian, grouping sets and the experimental results of the HeH$^+$ cation are presented in Appendix A.

\section{Experiments}

In order to efficiently implement VQE using entanglement measurements without the experimental costs of additional CNOT operations, we implement a photonic VQE using single-photon polarization and path degrees of freedom. Our photonic VQE presents four-dimensional quantum states or equivalently two qubits. With this encoding, a CNOT operation can be deterministically implemented via a polarizing beamsplitter. Therefore, our photonic QPU can clearly show the experimental advantage of introducing entanglement measurements in VQE.

\subsection{Experimental setup}

\begin{figure*}[b]
\centering
\includegraphics[width=4in]{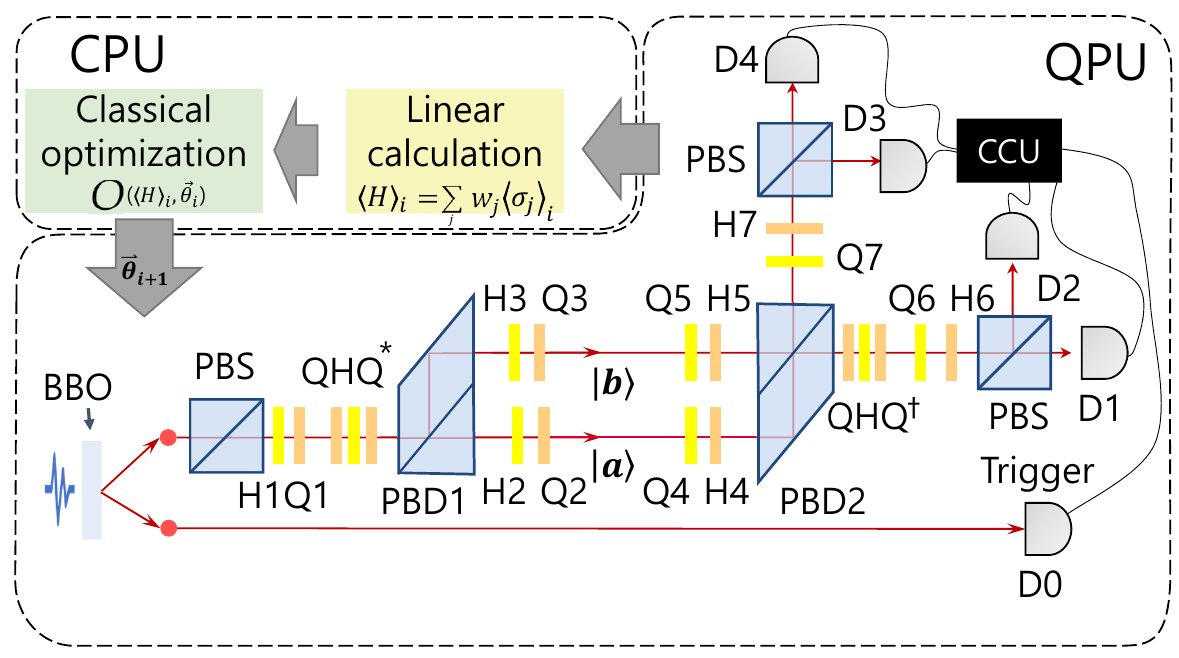}
\caption{Experimental setups for photonic VQE using paths and polarization of a single-photon. H: half waveplate, Q: quarter waveplate, PBD : laterally displaced polarizing beamdisplacer, PBS: polarizing beamsplitter, D: single-photon detector, CCU: coincidence counting unit.}
\label{vqe_setup}
\end{figure*}

Figure~\ref{vqe_setup} shows our photonic VQE experimental setup. We generate heralded single-photons via type-II spontaneous parametric down-conversion at a beta-barium borate (BBO) crystal pumped by femtosecond laser pulses at central wavelength 390~nm. One of the photon pairs is triggered by a single photon detector (D0), and the other is sent to the photonic VQE setup. The single-photon passes through a polarizing beamsplitter (PBS) to setup the initial polarization state. 

We utilize path ($|a\rangle$ and $|b\rangle$) and polarization ($|H\rangle$ and $|V\rangle$) degrees of freedom to encode quantum states into a single-photon state. The path degree of freedom is encoded by waveplates H1 (half-wave plate) and Q1 (quarter-wave plate) followed by a laterally displaced polarizing beam displacer (PBD1) which laterally splits horizontal and vertical polarization photons. Therefore, the amplitude and phase of optical paths can be controlled by the polarization state before PBD1. Then, we can control the polarization states of each path using waveplates (H2, Q2, H3 and Q3). Thus, we can prepare an ansatz state in terms of arbitrary two-qubit quantum state of $|\psi(\vec{\theta})\rangle=\alpha|aH\rangle+\beta|aV\rangle+\gamma|bH\rangle+\delta|bV\rangle$ which is parameterized by angles $\vec{\theta}$ of six waveplates. Note that QHQ$^*$ waveplates before PBD1 are utilized to compensate the relative phase drift between two optical paths. The relative phase of the two paths is monitored and compensated in every 10~min.  

The measurement of quantum states in this photonic QPU can be implemented in a reverse way of quantum state preparation, i.e., path modes $|a\rangle$ and $|b\rangle$ interfere at the other polarizing beam displacer (PBD2). In order to fully implement arbitrary projective measurement, the polarization states before and after PBD2 should be controlled by sets of waveplates \{H4, Q4, H5, Q5, H6, Q6, H7, Q7\}. Then, the single-photon is found at one of four outputs according to the projective measurement results. Note that sets of QHQ$^\dag$ are used to compensate the relative phase between two outputs of PBD2. The single photon detection at D1$\sim$D4 and two-fold coincidences with D0 are registered by a homemade coincidence counting unit (CCU)~\cite{park2015,park2021}.

We remark that our measurement setup can perform not only projective measurements on Pauli bases of each qubit but also their entangled bases, e.g., Bell bases, which require a CNOT operation. The PBD2 serves as the essence of the CNOT operation since the polarization states of the input determine the optical paths. In particular, by setting the angles of waveplates \{H4, Q4, H5, Q5, H6, Q6, H7, Q7\} at \{45$^\circ$, 90$^\circ$, 45$^\circ$, 0$^\circ$, 22.5$^\circ$, 45$^\circ$, 22.5$^\circ$, 45$^\circ$\}, four Bell states $|\psi^\pm\rangle$ and $|\phi^\pm\rangle$ are registered at D1$\sim$D4, respectively.

After the two-qubit quantum states are prepared and measured in all the necessary bases, we can estimate the expectation value of Hamiltonian through weighted sum over all the Pauli string. Then, we find new parameters, i.e., angles of waveplates, to prepare a next ansatz state to reduce the expectation value. This process is done by a classical processing unit (CPU), and here we utilize COBYLA method to perform parameter optimization. The tolerance of the optimizer is set to be 0.01, which means the optimizer finishes its process when the expectation values change less than 1$\%$. 


\subsection{Experimental results}

\begin{figure*}[b]
\centering
\includegraphics[width=6.3in]{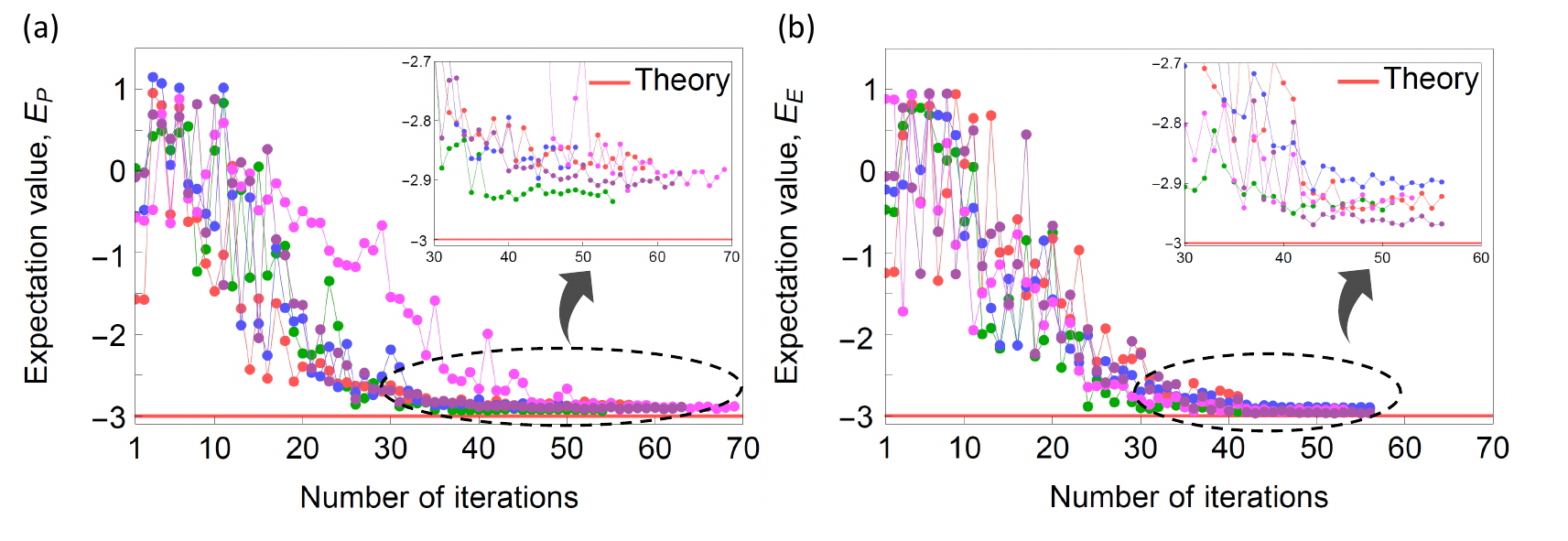}
\caption{VQE experimental results of $H_{2q}$ Hamiltonian with (a) Pauli measurements (VQE-P) and (b) entanglement measurements (VQE-E), respectively. $E_P$ and $E_E$ denote the expectation values from VQE-P and VQE-E, respectively. Five independent VQE runs are presented with different colors. In both cases, the expectation values $E_P$ and $E_E$ converges to around the theoretical value of $E_g=-3.000$. The tolerance of the optimizer is set to be 0.01.} 
\label{XXYYZZ}
\end{figure*}

We investigate the two-qubit antiferromagnetic Heisenberg Hamiltonian $H=XX+YY+ZZ$. We implement 5 times of VQE runs using Pauli measurement (VQE-P) and entanglement measurements (VQE-E), respectively. The VQE runs are independent each other and begin with a random initial ansatz state. Note that the total number of shots are fixed to be $N=9,000$ for both VQE methods. Since VQE-P requires three different measurement setups, these shots are equally shared by three different measurement setups of \{XX, YY, ZZ\} whereas VQE-E uses all the shots in a single Bell state measurement setup. We note that, although the total number of shots is identical for both VQE methods, the operation time of VQE-E is much shorter than that of VQE-P since it does not need to reconfigure the measurement setups, which is one of the major time delays in experiment.

Figure~\ref{XXYYZZ} presents the expectation values $E_P$ and $E_E$ from the optimization processes of (a) VQE-P and (b) VQE-E with respect to the number of iterations, respectively. It shows that both VQE-P and VQE-E can find the minimum eigenvalue of $E_g = -3.000$ with few tens of iterations. We summarize the number of iterations and estimated ground state energy in Table I to compare them with explicit numbers. It shows that the number of iterations is comparable within the standard deviation for VQE-P and VQE-E. However, the average of minimum expectation values of $E_P$ and $E_E$ present discrepancy. The average of $E_E$ for VQE-E is $1.33\%$ closer to the theoretical value than $E_P$ of VQE-P. The closer energy value achieved by VQE-E seems not to be due to the fewer shots used in each VQE-P setup, as the standard deviations of energies for both methods are nearly identical.  


\begin{table*}[b]
\centering
\begin{tabular}{|c|cc|cc|}
\hline
~~Methods~~ & \multicolumn{2}{c|}{VQE-P}              & \multicolumn{2}{c|}{VQE-E}               \\ \hline
Trials  & \multicolumn{1}{c|}{~Iteration~} &~~~~$E_P$~~~~  & \multicolumn{1}{c|}{~Iteration~} & ~~~~$E_E$~~~~    \\ \hline
1       & \multicolumn{1}{c|}{54}        & -2.931 & \multicolumn{1}{c|}{51}        & -2.942  \\ \hline
2       & \multicolumn{1}{c|}{59}        & -2.878 & \multicolumn{1}{c|}{56}        & -2.942  \\ \hline
3       & \multicolumn{1}{c|}{69}        & -2.904 & \multicolumn{1}{c|}{56}        & -2.906  \\ \hline
4       & \multicolumn{1}{c|}{63}        & -2.908 & \multicolumn{1}{c|}{53}        & -2.942  \\ \hline
5       & \multicolumn{1}{c|}{49}        & -2.879 & \multicolumn{1}{c|}{56}        & -2.969  \\ \hline\hline
Avg.    & \multicolumn{1}{c|}{58.8}      & -2.900   & \multicolumn{1}{c|}{54.4}      & -2.9402 \\ \hline
Stdev.   & \multicolumn{1}{c|}{7.76}      & 0.022  & \multicolumn{1}{c|}{2.30}      & 0.022   \\ \hline
\end{tabular}
\caption{VQE experimental results of $H_{2q}$ Hamiltonian with Pauli measurements (VQE-P) and entanglement measurements (VQE-E). Table of results for two-qubit Heisenberg model, $H_{2q}$. Iteration, $E_P$ and $E_E$ are the number of iterations and the minimum expectation value with VQE-P and VQE-E, respectively. Avg. and Stdev. denote the average and standard deviation of 5 times of VQE runs, respectively.}
\label{Heisen_table}
\end{table*}

We hypothesize that the discrepancy arises from practical experimental imperfections. To investigate this, we introduce an artificial error simulating practical imperfection of a controlled variable in the VQE processor. Specifically, we consider a scenario where the angles of the waveplates, which act as the QPU's parameters, have non-zero offset errors. We assume that the offset angles of waveplates follow the normal distribution as $f(\theta)=\frac{1}{\epsilon\sqrt{2\pi}}e^{-\frac{1}{2}(\frac{\theta-\theta_0}{\epsilon})^2}$ where $\theta_0$ is the errorless target angle and $\epsilon$ is the standard deviation of errors. By increasing $\epsilon$, we can simulate how increasing errors can affect the expectation values estimated by the VQE. Note that such experimental errors are common in usual photonic quantum information experiment~\cite{Lee2021prl}.
\begin{figure*}[t!]
\centering
\includegraphics[width=5.5in]{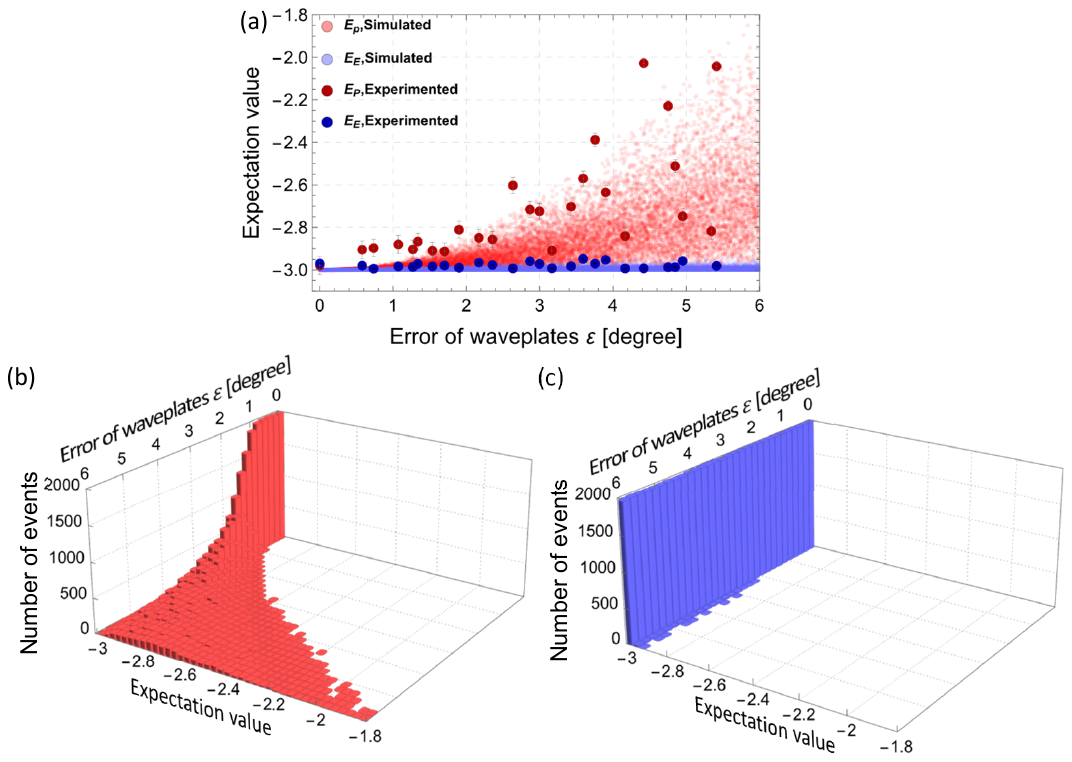}
\caption{VQE results of $H_{2q}$ Hamiltonian with increasing error of measurement apparatus, angles of waveplates. (a) The red and blue circles indicates the experimental results with VQE using Pauli and Bell measurements, respectively. The transparent circles are the Monte-Carlo simulation results for our VQE processes. The experimental values and error bars are obtained by averages and standard deviations of the five minimum values for a single VQE run. (b) and (c) show 3D histograms for the Monte-Carlo simulations of our VQE-P and VQE-E with respect to the error of waveplates, respectively.}
\label{errorXXYYZZ}
\end{figure*}

Figure~\ref{errorXXYYZZ} shows how the errors in waveplates for a measurement setup $\{$H4, Q4, H5, Q5, H6, Q6, H7, Q7$\}$ can cause VQE to malfunction. The dimmed red circles at Fig.~\ref{errorXXYYZZ}(a) show the Monte-Carlo simulation results of VQE-P ($E_P$), and their histogram is presented in Fig.~\ref{errorXXYYZZ}(b). As the errors increase, the attainable VQE expectation values $E_P$ diverges and can significantly deviates from the theoretical value. On the other hand, the Monte-Carlo simulation of VQE-E ($E_E$) almost always find the theoretical ground state energy $E_g=-3.000$ within the tolerance condition of optimizer, which are shown as dim blue circles in Fig.~\ref{errorXXYYZZ} (a) and the histogram of Fig.~\ref{errorXXYYZZ}(c). The experimentally obtained $E_P$ and $E_E$ for VQE-P and VQE-E are presented as red and blue circles in Fig.~\ref{errorXXYYZZ}(a), respectively. While VQE-P gives values that deviates from the theoretical value as $\sigma$ increases, VQE-E always find the results that almost match the theoretical minimum eigenvalue regardless of the presence of $\epsilon$. 

This error-robustness of VQE-E can be understood as follow. The impact of local errors which occur individually for distinct measurement setups of non-commuting observables, amplifies the error of expectation value of the Hamiltonian since all the expectations of Pauli strings would be appended together to obtain Hamiltonian expectation value. In contrast, as demonstrated in the Heisenberg model where all terms are measured in a single Bell measurement setup, all terms are affected by errors that occurred in the same system, and the heuristic optimization process guides the parameters towards a solution, regardless of the these errors. Therefore, for the antiferromagnetic Heisenberg Hamiltonian $H$, VQE using entanglement measurement (VQE-E) not only reduce the experimental cost by reducing the number of measurement setups but also provides the experimental error-robustness character.


\section{Conclusions}

We perform a VQE using a photonic quantum system, and its measurement-resource efficient realization by using entanglement measurements. Our four-dimensional quantum system, which has the same computational space as two-qubits, is determined by the polarizations and path degrees of freedom of a single-photon. We apply a deterministic CNOT operation using linear optical elements to implement a VQE using entanglement measurements without extra experimental burden. This shows that photonic quantum simulator based on multiple degrees of freedom of photons has the advantages of being efficient for entanglement measurements.

Our approach, which considers entanglement measurements, can reduce the number of measurement setups compared to the conventional VQE method that only uses Pauli measurements. Furthermore, we demonstrate that a VQE using entanglement measurements can robustly handle erroneous operations caused by the practical imprecisions in optics, i.e. non-zero offset angles of waveplates. We remark that the advantages of utilizing entanglement measurement in photonic VQE with multiple degrees of freedom would be maintained for larger number of photons and/or modes~\cite{erhard2020, yoo2018, wang2018, hong2020}. Thus, our work offers an experimentally resource-efficient approach to scaling up photonic quantum processors.

\section*{Acknowledgements}
\noindent This research was funded by Korea Institute of Science and Technology (2E32941, 2E32971), National Research Foundation of Korea (2023M3K5A1094805, 2022M3K4A1094774, 2022M3K4A1097119, 2022M3H3A106307411,2022M3E4A1043330).




\section*{References}

\bibliographystyle{iop}

\newpage
\appendix
\section{VQE with entanglement measurements for HeH$^+$ Hamiltonian}
\label{a:HeH}

\begin{figure*}[t]
\centering
\includegraphics[width=5.5in]{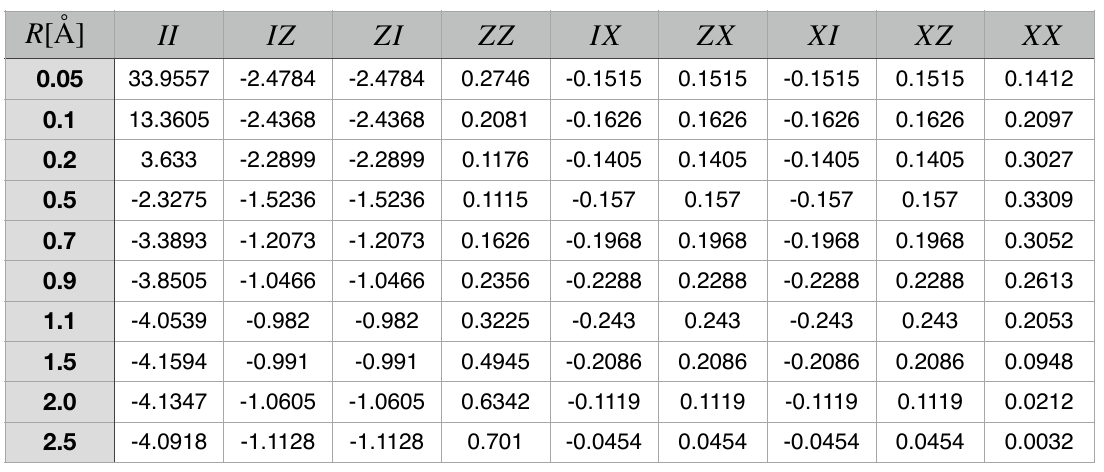}
\caption{The weights of Pauli strings of qubit Hamiltonian for Helium Hydride cation. The coefficients are represented by different values for different interatomic distances between helium and hydrogen atoms (See Eq.~(\ref{qHam})).}
\label{HeHHam}
\end{figure*}

In terms of the creation and annihilation operators, the problem Hamiltonian can be represented as
\begin{eqnarray}
H = \sum_{p,q} h_{p,q} a^\dagger_p a_q + \sum_{p,q,r,s} h_{p,q,r,s} a^\dagger_p a^\dagger_q a_r a_s + \cdots,
\label{Ham} \nonumber \\ 
\end{eqnarray}
where $a^\dagger$ and $a$ denote the creation and annihilation operators, respectively, together with interaction coefficients $h$. Through a transformation such as Jordan-Wigner or Bravyi-Kitaev transformation, Eq.~(\ref{Ham}) would be transformed into an weighted sum of Pauli strings,
\begin{eqnarray}
H = \sum_j w_j \sigma_j + \sum_{j,k} w_{j,k} \sigma_j \otimes \sigma_k+\cdots,
\label{qHam}
\end{eqnarray}
where the $\sigma_j$ is a Pauli operator with $j\in \{0,1,2,3\}$, which corresponds to $\{I, X, Y, Z \}$, respectively. The weights $w_j, w_{j,k},\cdots $ can be achieved in the transformation we choose. Here we consider Jordan-Wigner transformation, and show the Hamiltonian of HeH$^+$ with respect to the interatomic distance $R$, using Psi4 package~\cite{peruzzo2014}. The weights are listed in the Fig.~\ref{HeHHam}.

Two observables are simultaneously measurable if and only if they are commute. Therefore, we can measure commuting observables at the same time through the same measurement setup. Here, we consider two methods to consider commutativity, i.e., Qubit-wise Commutativity (QWC) and General Commutativity (GC), with GC being the most general method that covers QWC. Two Pauli strings satisfy qubit-wise commutativity if at each index, the corresponding two Pauli operators commute. It is also called TPB since the simultaneous eigenbasis can be expressed as a tensor product across each qubit index without entanglement. Note that grouping Pauli strings into QWC context in optimal is known as a NP-hard problem~\cite{Gokhale2019}. 

QWC is sufficient but not necessary for commutativity between Pauli strings. General commutativity (GC) is an extension that considering commutativity between whole Pauli strings. As an example, a set $\{XX, YY, ZZ\}$ is a commuting family even non of the pairs are QWC. 
Here, rather than considering the most general one, we consider only the commutativity between length-2 Pauli strings (e.g., $XX$, $YY$, $ZZ$) together with QWC, taking account the utilization of Bell measurement. HeH$^+$ Hamiltonian has 9 Pauli strings in terms of qubit Hamiltonian, $\{II, IZ, ZI, ZZ, IX, ZX, XI, XZ, XX \}$. Here we make three groups of Pauli strings $\{XX, ZZ, II \}$, $\{XI, XZ, IZ \}$ and $\{IX, ZI, ZX \}$. The elements of first group $\{XX, ZZ, II \}$ are not commute each other in the viewpoint of QWC. But, by considering GC, they are commute and simultaneously measurable in terms of Bell basis measurements, as shown in Eq.~(\ref{byBSM}) in main text. And other two groups are grouped by QWC, therefore we can implement VQE for HeH$^+$ by realizing three measurement setups, one for each group. 

The grouping method reduces the costs for state measurements in VQE compared to the case where all the Pauli strings with 9 distinct setups are considered, and is also smaller than the case where only QWC with four groups is dealt with, specifically, $\{IX,ZX \}, \{XI,XZ \}, \{XX\}$ and $\{II,IZ,ZI,ZZ \}$.

\begin{figure*}[t]
\centering
\includegraphics[width=2.9in]{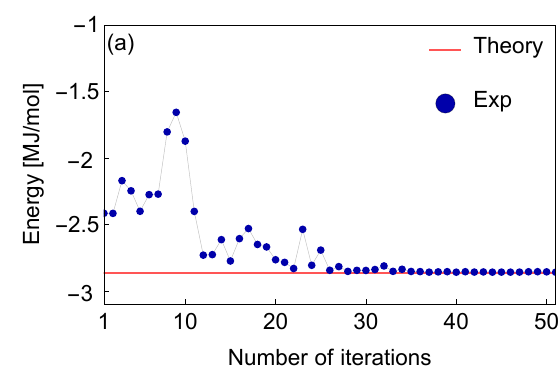}
\includegraphics[width=2.9in]{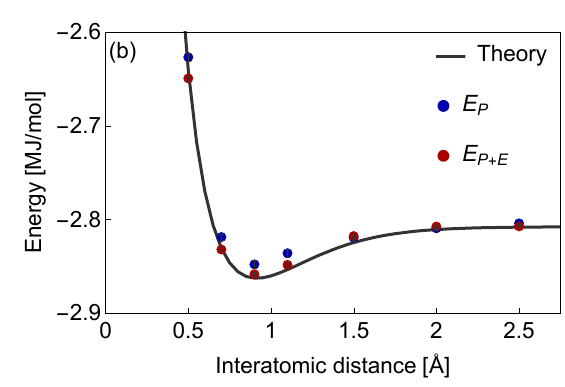}
\caption{Results of VQE experiment using entanglement measurements. Hamiltonian of HeH$^{+}$ cation is considered. (a) for $R=0.9$~\AA, the convergence of the estimated energy value as the VQE process with Pauli+entangled (P+E) measurements progresses (b) the ground state energy with respect to the interatomic distance between He and H$^+$. The blue and red circles are the VQE results with local Pauli measurements solely ($E_P$) and Pauli+entanglement measurements ($E_{P+E}$), respectively. Each point is obtained by averaging the five smallest values after minimization, and their standard deviations are sufficiently small to fit within the points.}
\label{He-H}
\end{figure*}

Figure~\ref{He-H} shows the experimental results of the ground state energy of HeH$^+$ cation using our photonic VQE. Here, we denote the cases as the case of Pauli$+$entanglement measurements (P+E) and the case of Pauli measurements only (P). Figure~\ref{He-H}(a) shows the procedure of VQE for HeH$^+$ cation Hamiltonian using P+E at the interatomic distance $R=0.9$~\AA~with respect to the number of iterations. The expectation value converges to the theoretical energy value effectively after about 30 iterations. Figure~\ref{He-H}(b) shows the ground state energy of HeH$^+$ cation as a function of interatomic distance between He and H$^+$. The blue and red circles indicate experimental energy values obtained by VQE using Pauli measurements alone ($E_P$) and Pauli$+$entanglement measurements ($E_{P+E}$), respectively. While both VQE results present good agreement with the theoretical values (black curve), $E_{P+E}$ shows slightly closer to the theoretical values. For instance at $R=0.9$~\AA, the theoretical ground state energy is $E_{th} = -2.863$ MJ/mol, and the VQE finds ground state energy of $E_{P} = -2.848\pm 0.004$ MJ/mol and $E_{P+E} = -2.858 \pm 0.002$ MJ/mol, respectively. In conclusion, we confirmed that both VQE with Pauli measurement alone and Pauli$+$entanglement measurement find the ground energy well, the later can be implemented with fewer measurement setups.

\end{document}